\documentclass[a4paper,10pt,twoside]{cpc-hepnp}

\usepackage{multicol}
\usepackage{graphicx}
\usepackage{booktabs}
\usepackage{amssymb,bm,mathrsfs,bbm,amscd}
\usepackage[tbtags]{amsmath}
\usepackage{lastpage}
\usepackage{CJK}
\usepackage{color}

\def\aap{Astron.\ Astrophys.}
\def\apj{Astrophys.\ J.}
\def\apjl{Astrophys.\ J.\ Lett.}
\def\apjs{Astrophys.\ J.\ Supp.}
\def\aj{Astron.\ J.}
\def\mnras{Mon.\ Not.\ Roy.\ Astron.\ Soc.}
\def\physrep{Phys.\ Rept.}
\def\prd{Phys.\ Rev.\ D}
\def\prl{Phys.\ Rev.\ Lett.}

\def\araa{Annu.\ Rev.\ Astron.\ Astrophys.}
\def\jcap{J.\ Cosmol.\ Astropart.\ Phys.}

\begin{document}
\begin{CJK*}{GB}{gbsn}


\fancyhead[c]{\small } \fancyfoot[C]{\small 010201-\thepage}


\title{Searching for Gamma-Ray Emission from Reticulum II by Fermi-LAT}

\author{
      Yi Zhao $^{1,2}$
\quad Xiao-Jun Bi $^{3,4}$
\quad Peng-Fei Yin $^{3}$
\quad Xin-Min Zhang $^{1,2,4}$
}

\maketitle

\address{
$^1$ Theoretical Physics Division, Institute of High Energy Physics,\\
Chinese Academy of Sciences, Beijing 100049, China\\

$^2$ Theoretical Physics Center for Science Facilities, \\
Chinese Academy of Sciences, Beijing 100049, China \\

$^3$ Key Laboratory of Particle Astrophysics, Institute of High Energy Physics,\\
Chinese Academy of Sciences, Beijing 100049, China\\

$^4$ School of Physical Sciences, University of Chinese Academy of Sciences, Beijing 100049, China
}

\begin{abstract}
Recently, many new dwarf spheroidal satellites (dSphs) have been discovered by
the Dark Energy Survey (DES). These dSphs are
ideal candidates for probing for gamma-ray emissions from dark matter (DM) annihilation.
However, no significant signature has been found by the Fermi-LAT dSph observations. In this work, we reanalyze the Fermi-LAT Pass 8 data from the direction of Reticulum II, where a slight excess has been reported by some previous studies.
We treat Reticulum II (DES J0335.6-5403) as a spatially extended source, and find that no significant gamma-ray signature is observed. Based on this result, we set upper-limits on the DM annihilation cross section.
\end{abstract}


\begin{pacs}
95.35.+d, 95.85.Pw, 98.52.Wz   \ \ \ \ \ \ \   \textbf{DOI:} 10.1088/1674-1137/42/2/025102
\end{pacs}


\begin{multicols}{2}

\section{Introduction}

Many astrophysical and cosmological observations have shown that cold dark matter (DM) particles
make up $25.8\%$ of the Universe, while baryons only make up $4.8\%$ \cite{2016A&A...594A...1P}. Weakly interacting massive particles (WIMPs) are
a popular candidate for cold DM \cite{1996PhR...267..195J,2000RPPh...63..793B,2005PhR...405..279B}.
WIMPs can annihilate or decay to standard model particles, such as electrons/positrons, photons, neutrinos, etc.
These processes provide an indirect way to probe the DM properties.
Compared with charged particles, photons are not affected by the Galaxy's magnetic field and suffer from
less energy loss in the propagation process. Therefore, probing gamma-rays is a very promising approach for indirect DM  detection.

DM annihilation can directly produce photons at loop level, or indirectly produce photons through cascade decays,
final state radiation, and inverse Compton scattering processes. It is possible to observe $\gamma$-rays from regions with high DM densities  using high precision detectors, such as the space-borne $\gamma$-ray detector Fermi Large Area Telescope (Fermi-LAT) \cite{2009ApJ...697.1071A}.
Many studies have been performed to probe the DM signatures in the Fermi-LAT data from large astrophysical systems, such as the Galactic halo \cite{2011PhRvD..84l3005H,2012ApJ...761...91A,2012PhRvD..86h3511A,
2010PhRvL.104i1302A,2012PhRvD..86b2002A,2012JCAP...08..007W,2013PhRvD..88h2002A}, galaxy clusters \cite{2010JCAP...05..025A},
and Galactic DM substructures \cite{2012A&A...538A..93Z,2012ApJ...747..121A,2012JCAP...11..050Z}.
However, the contribution of DM annihilation to the reported GeV $\gamma-$ rays has not been confirmed \cite{2009arXiv0910.2998G,2009arXiv0912.3828V,2011PhLB..697..412H,
2011PhRvD..83h3517H,2012PhRvD..86h3511A,2013PhRvD..88h3521G,
2013PDU.....2..118H,2016PDU....12....1D,2015JCAP...03..038C,
2015PhRvD..91f3003C,2015PhRvD..91l3010Z,
2015JCAP...12..056G,2016PhRvL.116e1102B,2016PhRvL.116e1103L}.

Nearby dwarf spheroidal satellites (dSphs) are
another promising candidate for detecting gamma-ray emission from DM annihilation, due to the lack of astrophysical $\gamma$-ray sources \cite{1998ARA&A..36..435M,2009ApJ...696..385G}.
Recently, more than twenty dSphs have been discovered by the Sloan Digital Sky Survey (SDSS) \cite{2000AJ....120.1579Y},
which has a deep and systematic coverage of the northern celestial hemisphere.
Numerous studies have been performed to search for the $\gamma$-ray emission from these dSphs in the Fermi-LAT data,
but no significant signal has been found \cite{2010ApJ...712..147A,2011PhRvL.107x1302A,2011PhRvL.107x1303G,
2012APh....37...26M,2014PhRvD..89d2001A,2015PhRvD..91h3535G,2015PhRvL.115w1301A}. The ongoing Dark Energy Survey (DES) \cite{2005astro.ph.10346T,2016MNRAS.460.1270D}, which is an optical instrument working on the southern hemisphere,
has also reported $16$ new dSphs in $2015$ \cite{2015ApJ...807...50B,2015ApJ...805..130K,2015ApJ...813..109D}.
Among these new candidates, there seem to be slight excesses at GeV level in the directions of Reticulum II (DES J0335.6-5403)
and Tuc III (DES J2356-5935) \cite{2015PhRvL.115h1101G,2015JCAP...09..016H,2015arXiv151109252L}.
However, the excess of Reticulum II from the the latest analyses of Fermi-LAT Pass 8 data is less significant than that from the previous analyses of Pass 7 data. \cite{2015ApJ...809L...4D,2017ApJ...834..110A}.
This difference may be caused by the improvement of the data, as declared by the Fermi-LAT collaboration \cite{2015ApJ...809L...4D}.

Due to the lack of  information about the spatial extension of DM halos, the newly discovered dSphs have usually been treated as point-like sources in previous gamma-ray searches. In Ref. \cite{2015ApJ...808L..36B}, the authors analyzed the density profile of Reticulum II, and calculated
the $J$ factor for DM annihilation within some typical extended angles. By using these results, we search for $\gamma$-ray emission in the Fermi-LAT Pass 8 data from the direction of Reticulum II, which is taken to be an extended source. No significant signature has been found in our analysis. We set constraints on the DM annihilation cross section for several DM annihilation channels.

This paper is organized as follows. In Section 2, we analyze the Fermi-LAT data in the direction of
Reticulum II for some typical spatially extended angles and set upper-limits on the $\gamma$-ray flux.
In Section 3, we set constraints on the DM annihilation cross section by the Fermi-LAT $\gamma$-ray observation of Reticulum II,
and discuss the main element which may influence the results.
Section 4 gives our conclusion.

\section{Data analysis}
In our study, we use Science Tools version v10r0p5 to analyse approximately 7 years of Fermi-LAT data, recorded from
August 2008 to June 2015, with the Pass 8 photon data selection. Events from the Pass 8 SOURCE-class
at energies between 500 MeV and 500 GeV are used. In order to reduce the $\gamma$-ray
contamination from the earth's limb, events with zenith angles larger than $100^{\circ}$ have been rejected,
and the recommended filter cut $(DATA\_QUAL>0, LAT\_CONFIG==1)$ is applied. We create a $10^{\circ} \times 10^{\circ}$
square region as the ROI around the dSph center with pixels of $0.1^{\circ}$, and consider 8 logarithmic energy bins from
500 MeV to 500 GeV. We use the Galactic $\gamma$-ray diffuse model $gll\_iem\_v06.fit$ and
the isotropic extragalactic $\gamma$-ray diffuse spectrum $iso\_P8R2\_SOURCE\_V6\_v06.txt$ as the diffuse background.
The third LAT source catalog (3FGL) \cite{2015ApJS..218...23A} is used to deal with the known point sources. In the analysis,
we carry out a global fit in the entire energy range and then fix all backgrounds (both point-like and diffuse backgrounds)
in each energy bin. The instrument response functions (IRFs) $P8R2\_SOURCE\_V6$ are taken according to the above
data selection.

In each energy bin the putative $\gamma$-ray signal from dSph Reticulum II is modeled as a power-law
($dN/dE \propto E^{-\Gamma}$) with a spectral index of $\Gamma=2$. The spatial map for the extended source
is indispensable, and is modeled by the DM density profile here.
The $J$ factor is defined as the integration of the squared DM density along the line-of-sight over the solid angle
\begin{eqnarray}\label{eq:zy1}
J = \int J_{\Omega} d\Omega =\iint \rho^{2}(l,\Omega) dld\Omega.
\end{eqnarray}
The solid angle $\Delta \Omega$ can be expressed as $\Delta \Omega = 2\pi \times [1-cos(\alpha_{int})]$,
where $\alpha_{int}$ is the integral angle.
In Ref. \cite{2015ApJ...808L..36B}, the DM density profile of Reticulum II is reconstructed by performing an optimized spherical
Jeans analysis for the kinematic data obtained from the Michigan/Magellan Fiber System. The authors provided the
$J$ factors for different integral angles $\alpha_{int}$, of which the typical values are listed
in Table \ref{tab:zy1} \cite{2015ApJ...808L..36B}. Although these values are derived by the choice of the Einasto profile, the choice of other DM profiles, such as the NFW profile, would result in very similar $J$ factors in the same procedure of the Jeans analysis ~\cite{2015MNRAS.446.3002B}. Therefore, such a choice would not significantly impact our final results.

\begin{center}
\tabcaption{$J$ factors for Reticulum II.}
 \begin{tabular}{lcc}
  \hline\hline\noalign{\smallskip}
   $\alpha_{int}$  &\qquad\qquad\qquad    & $\log_{10}(J(\alpha_{int}))$  \\
   $(deg)$         &    & $(\log_{10}[GeV^{2}cm^{-5}])$   \\
  \hline\noalign{\smallskip}
  0.1     &    & $18.8^{+0.6}_{-0.5}$      \\
  0.5     &    & $19.6^{+1.0}_{-0.7}$      \\
  1       &    & $19.8^{+1.2}_{-0.9}$      \\
  \noalign{\smallskip}\hline\hline
\end{tabular}
\label{tab:zy1}
\end{center}

By using these results,
we show the relation between $\alpha$ and $J_{\Omega}$ in Fig. \ref{fig:zy1}, which represents the information of the spatial extension.
Note that $J_{\Omega}$ is an un-integrated astrophysical factor
within an solid angle $\Delta \Omega$, as denoted in Eq. \ref{eq:zy1}.
Then we can construct three spatial maps of
Reticulum II for three typical extended angles of
$0.1^{\circ}$, $0.5^{\circ}$, and $1^{\circ}$.

\begin{center}
\includegraphics[width=0.9\columnwidth, angle=0]{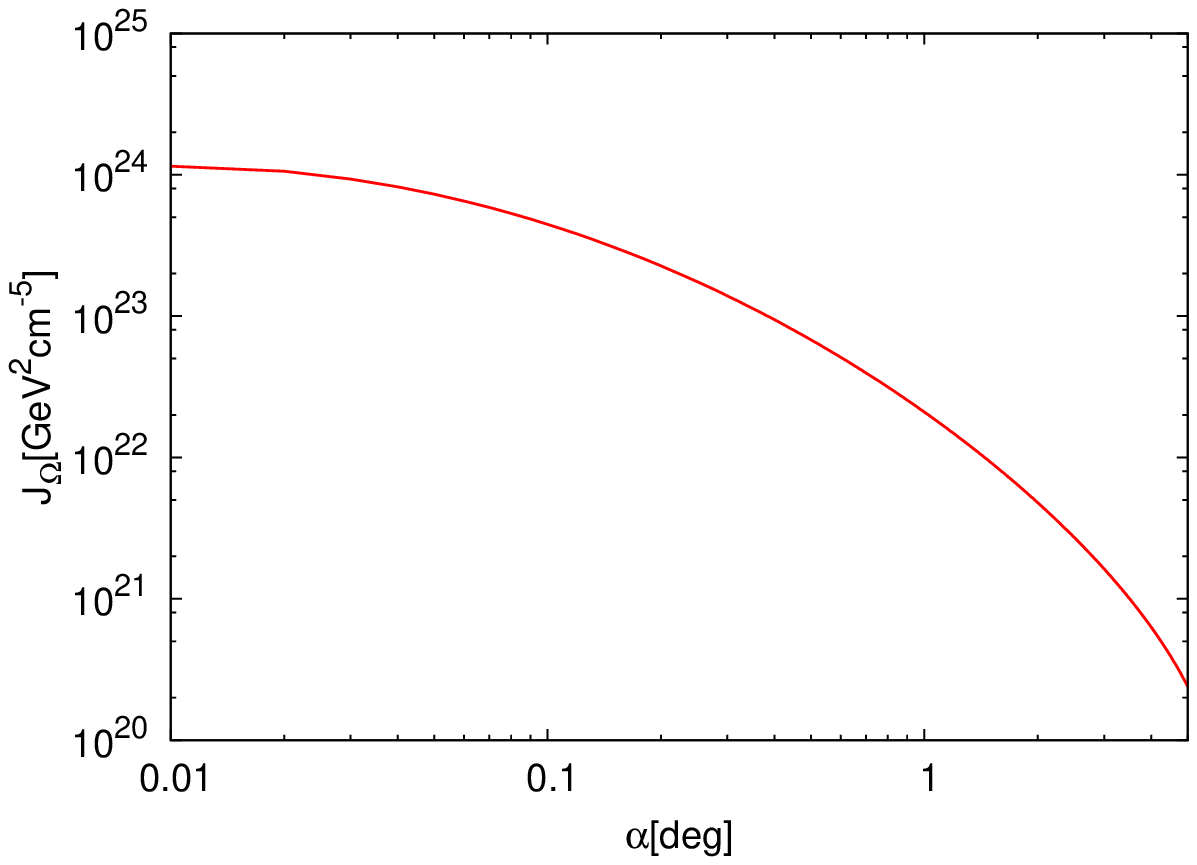}
\figcaption{The un-integrated astrophysical factor of Reticulum II as a function of the angle.}
\label{fig:zy1}
\end{center}

\begin{figure*}[!htb]
\centering
\includegraphics[width=0.9\columnwidth, angle=0]{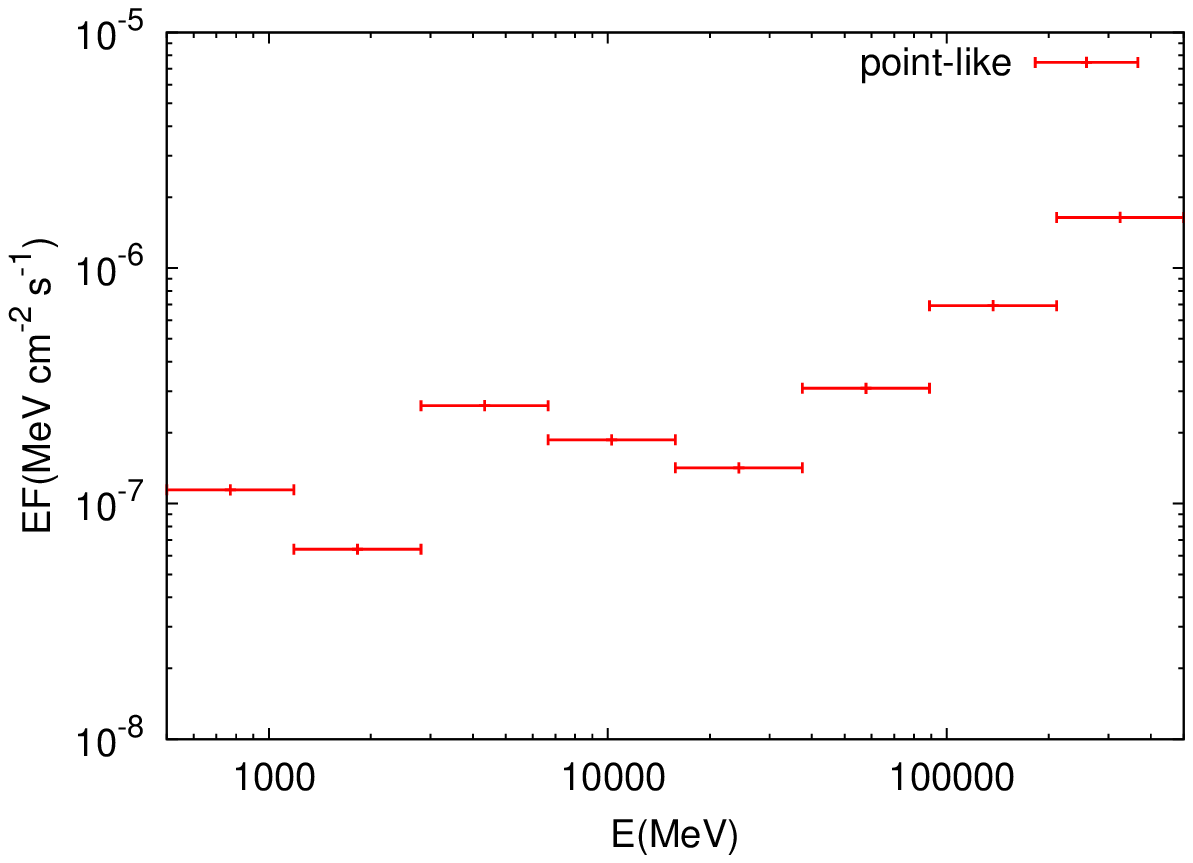}
\includegraphics[width=0.9\columnwidth, angle=0]{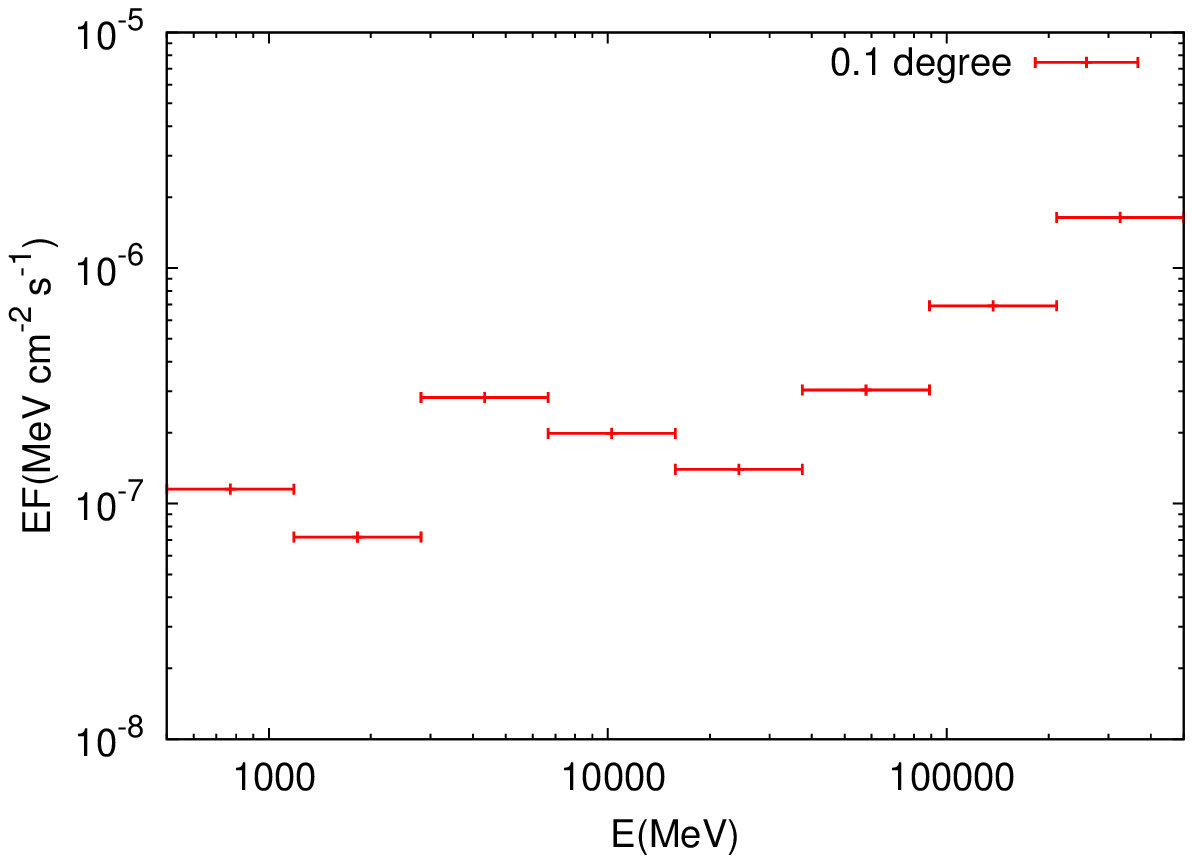}
\includegraphics[width=0.9\columnwidth, angle=0]{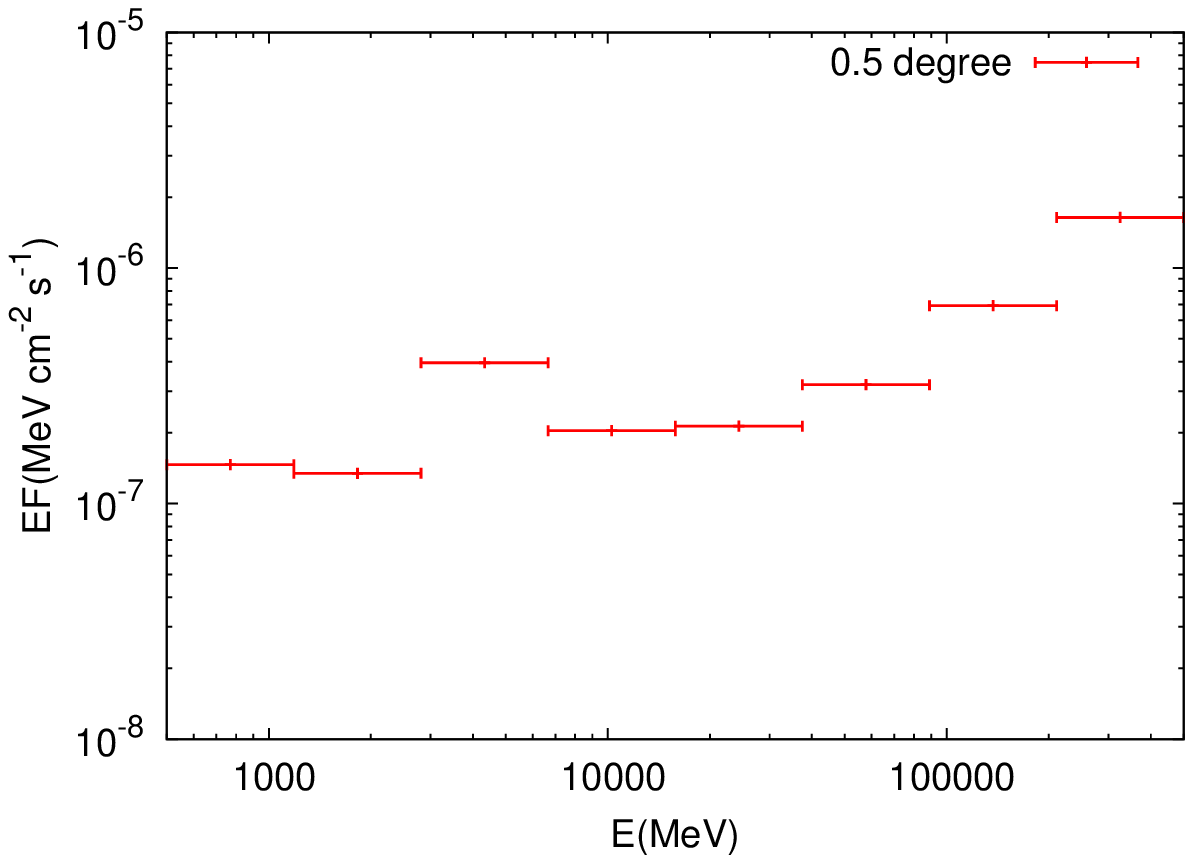}
\includegraphics[width=0.9\columnwidth, angle=0]{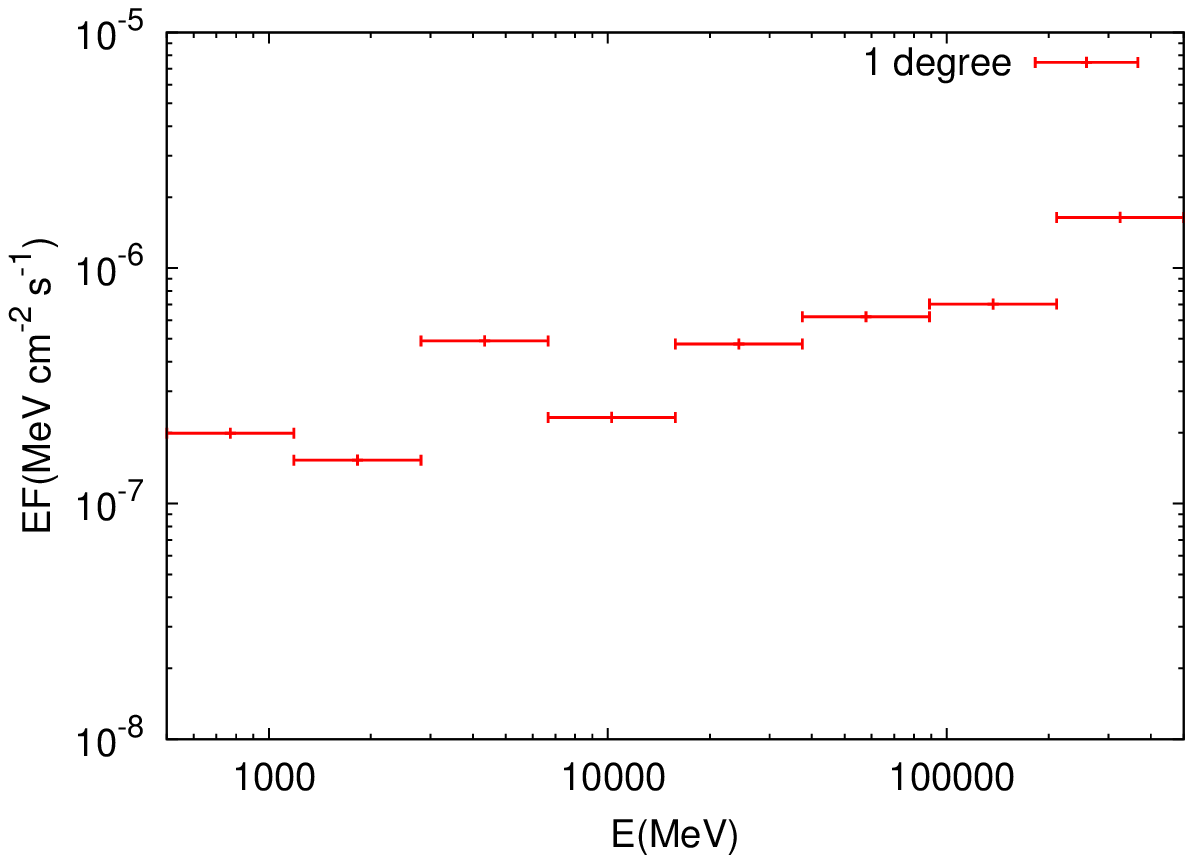}
\figcaption{The upper limits on the $\gamma$-ray flux of Reticulum II at $95\%$ CL for the point-like assumption and spatially extended source assumption for three extended angles.}
\label{fig:zy2}
\end{figure*}

Following the procedure of Ref. \cite{2016PhRvD..93h3513Z}, we fit the $\gamma$-ray signature in each energy bin
separately and then derive limits on the $\gamma$-ray flux, which are independent of the specific spectral model.
We set a range of $\gamma$-ray flux in each energy bin, and are then able to get the corresponding likelihood for each flux.
By using this flux and likelihood information, we are able to get the upper limits on the $\gamma$-ray flux at
$95\%$ confidence level (CL).
We consider the cases where
Reticulum II is modeled as a point-like source or an extended source with three extended angles, and show the corresponding upper-limits on the $\gamma$-ray flux at $95\%$ confidence level (CL) in Fig. \ref{fig:zy2}. It can be seen that there seems to be a slight $\gamma$-ray excess in the third energy bin, between 2.8 GeV and 6.7 GeV.

The test statistic (TS) value of the $\gamma$-ray signature is given by
\begin{eqnarray}\label{eq:zy1}
TS=-2\log \left(\frac{\mathcal{L}_{max,0}}{\mathcal{L}_{max,1}}\right),
\end{eqnarray}
where $\mathcal{L}_{max,1}$ and $\mathcal{L}_{max,0}$ are the maximal likelihoods of the cases with and without exotic signals, respectively.
We consider a set of DM masses from 2 GeV to 10 TeV
for six DM annihilation channels (when the final states are kinematically allowed), namely $e^+e^-$, $\mu^+\mu^-$, $\tau^+\tau^-$, $u\bar{u}$, $b\bar{b}$, and $W^+W^-$.
We find that an annihilating DM with a mass of 25 GeV to $\tau^+\tau^-$ final states contributes the most
significant excess, while the corresponding significance is only about $2\sigma$. The significances of the other channels are smaller than that of the $\tau^+\tau^-$ channel. For example, the best TS values in the point-like source case for the $e^+e^-$, $\mu^+\mu^-$, $u \bar{u}$, and $b\bar{b}$ channels are 3.73, 5.34, 4.28, and 4.43, respectively.

Although different extended angles result in different TS values, as listed in Table \ref{tab:zy2}, these values do not vary very much. This result indicates that the extended angle of Reticulum II does not have a significant impact on the TS value. We show the residuals of a $2^{\circ} \times 2^{\circ}$  ROI centered on Reticulum II in Fig. \ref{fig:zy3},
which has been smoothed with a $0.2^{\circ}$ Gaussian kernel. Although it is difficult to recognize any obvious structure from this residual map, the residuals are not uniformly distributed. This means that the variation of the TS value with the extended angles may be affected by fluctuations.

\begin{center}
\tabcaption{TS values for different spatial extensions.}
 \begin{tabular}{lcc}
  \hline\hline\noalign{\smallskip}
   Spatial type  &\qquad\qquad\qquad\qquad    & TS value  \\
  \hline\noalign{\smallskip}
  Point-like            &   & 5.46     \\
  Extend $0.1^{\circ}$  &   & 5.96     \\
  Extend $0.5^{\circ}$  &   & 6.46      \\
  Extend $1^{\circ}$    &   & 4.85      \\
  \noalign{\smallskip}\hline\hline
\end{tabular}
\label{tab:zy2}
\end{center}

\begin{center}
\includegraphics[width=0.9\columnwidth, angle=0]{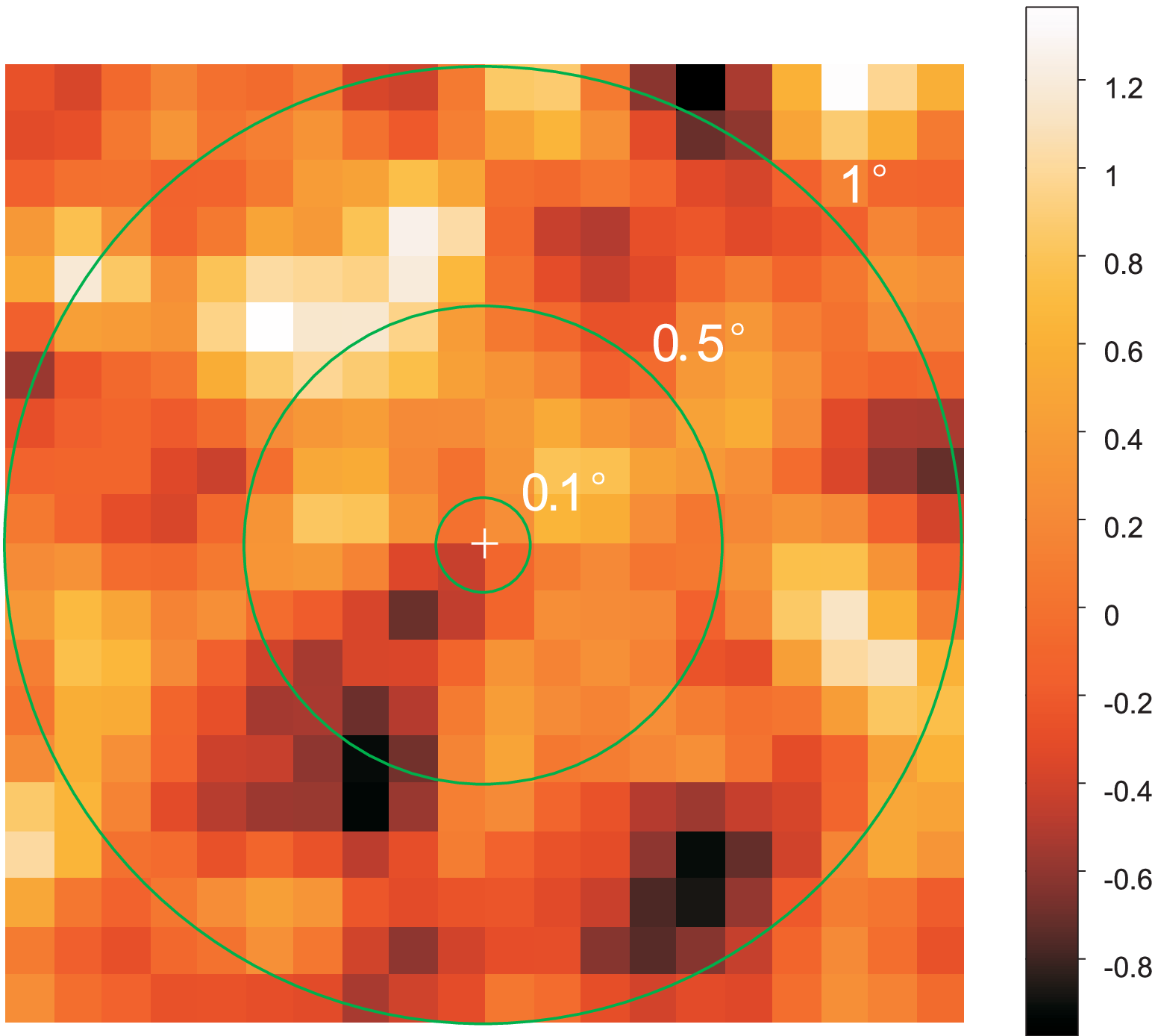}
\figcaption{The residual map of a $2^{\circ} \times 2^{\circ}$ ROI centered on Reticulum II, smoothed with a $0.2^{\circ}$ Gaussian kernel.}
\label{fig:zy3}
\end{center}

Note that our results are consistent with those given by the Fermi-LAT collaboration based on the Pass 8 data. In Refs.~\cite{2015ApJ...809L...4D,2017ApJ...834..110A}, only a slight excess from Reticulum II with a significance smaller than $2.5\sigma$ (TS$=6.8\sim7.0$) for the $\tau^+\tau^-$ channel was reported. However, Refs. \cite{2015PhRvL.115h1101G,2015JCAP...09..016H} found a significant $\gamma$-ray excess with a significance larger than $3 \sigma$ in the Pass 7 data. The main difference may be caused by the different versions of the data, as explained by the Fermi-LAT collaboration. For instance, compared with the Pass 7 data, the Pass 8 data improves the point source sensitivity by a factor of $30\%-50\%$ \cite{2015ApJ...809L...4D}.

\section{Constraints on the DM annihilation cross section}

In this section, we set upper limits on the DM annihilation cross section from the Fermi-LAT $\gamma$-ray observation of Reticulum II.
The expected integrated $\gamma$-ray flux from DM annihilation
is given by
\begin{eqnarray}\label{eq:zy2}
\Phi(E) = \frac{\left<\sigma v\right>}{8\pi m_{DM}^2}\times \int \frac{dN_\gamma}{dE_\gamma} dE_\gamma \times J ,
\end{eqnarray}
where $\left<\sigma v\right>$ is the thermally averaged DM annihilation cross section,
$m_{DM}$ is the DM mass, $\frac{dN_\gamma}{dE_\gamma}$ is the differential
$\gamma$-ray spectrum from one DM pair annihilation, and $J$ is the astrophysical factor
expressed as Eq. \ref{eq:zy1}.
Note that $\frac{dN_\gamma}{dE_\gamma}$ should be a sum of the photons from all possible DM
annihilation final states according to the detailed DM model. Here we only consider the $\gamma$-ray
contribution from a sole annihilation channel, including $e^+e^-$, $\mu^+\mu^-$, $\tau^+\tau^-$, $u\bar{u}$, $b\bar{b}$, and $W^+W^-$, by using the initial spectra given by PPPC \cite{2011JCAP...03..051C,2011JCAP...03..019C}.
Note that more stringent constraints should be derived
from a combined analysis including all the dSphs, or at least including those with large $J$ factors.
Here we attempt to find the impact of varying spatially extended angles of Reticulum on TS values and constraints, and hence only
consider the Reticulum result in our analysis.

\begin{figure*}[!htb]
\centering
\includegraphics[width=0.9\columnwidth, angle=0]{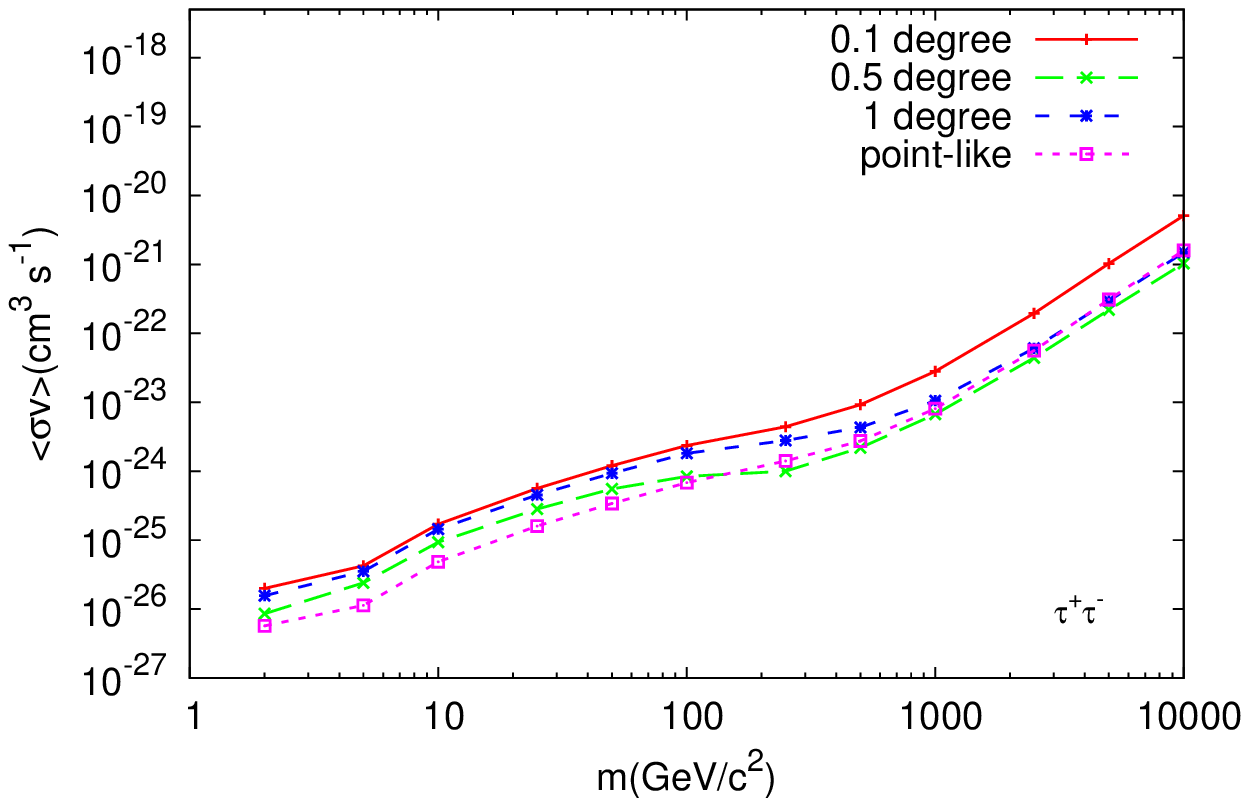}
\includegraphics[width=0.9\columnwidth, angle=0]{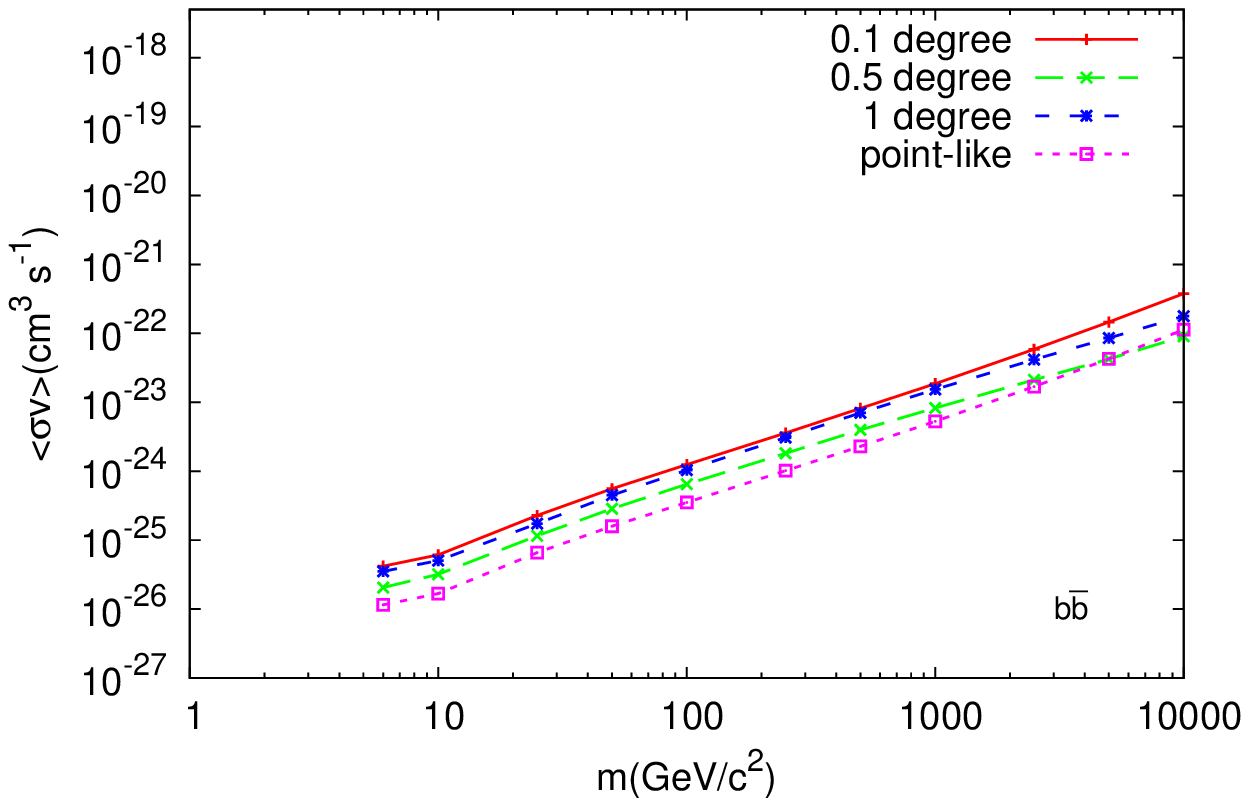}
\includegraphics[width=0.9\columnwidth, angle=0]{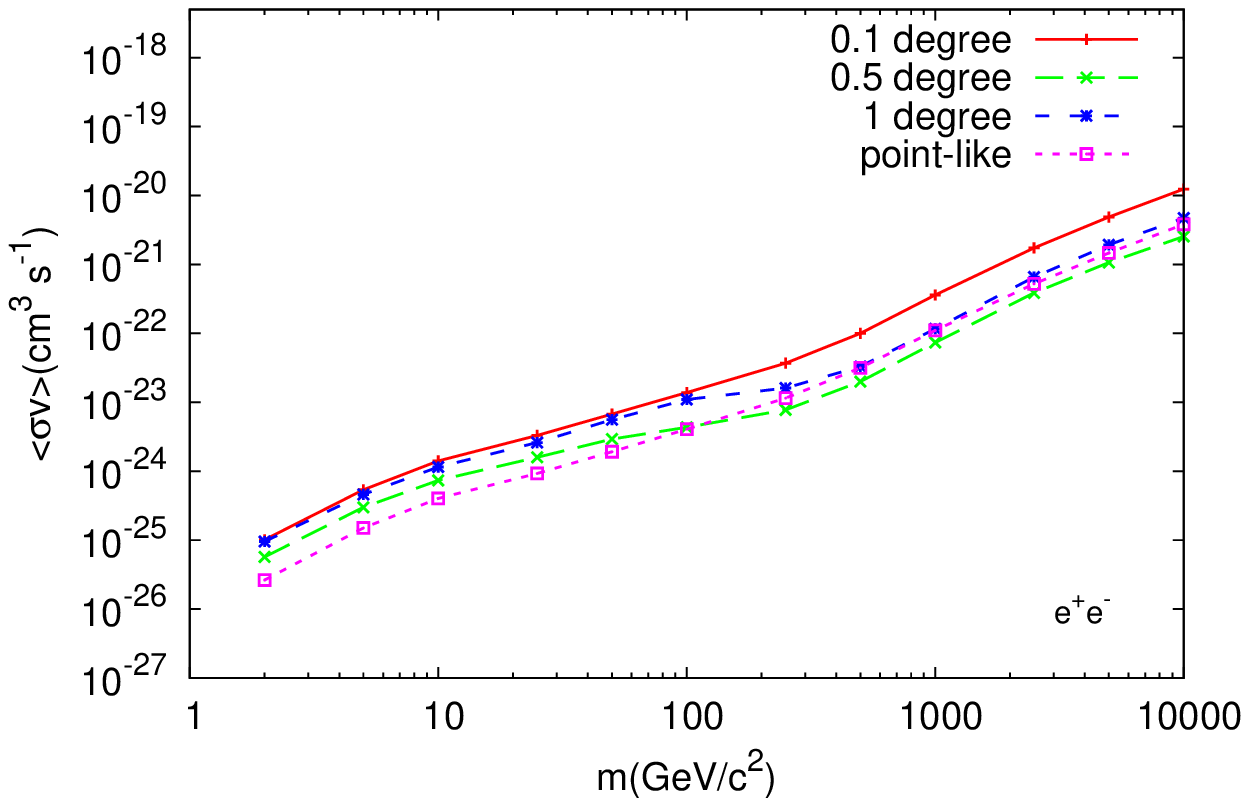}
\includegraphics[width=0.9\columnwidth, angle=0]{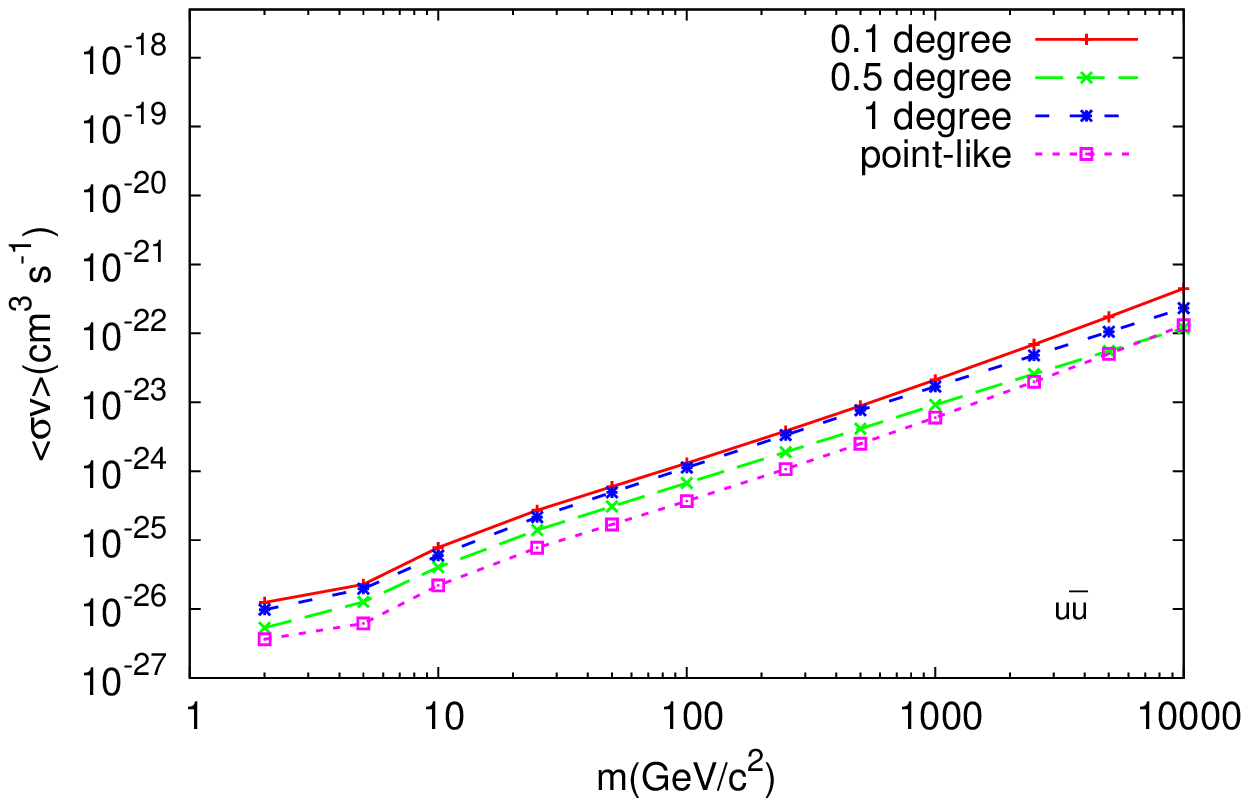}
\includegraphics[width=0.9\columnwidth, angle=0]{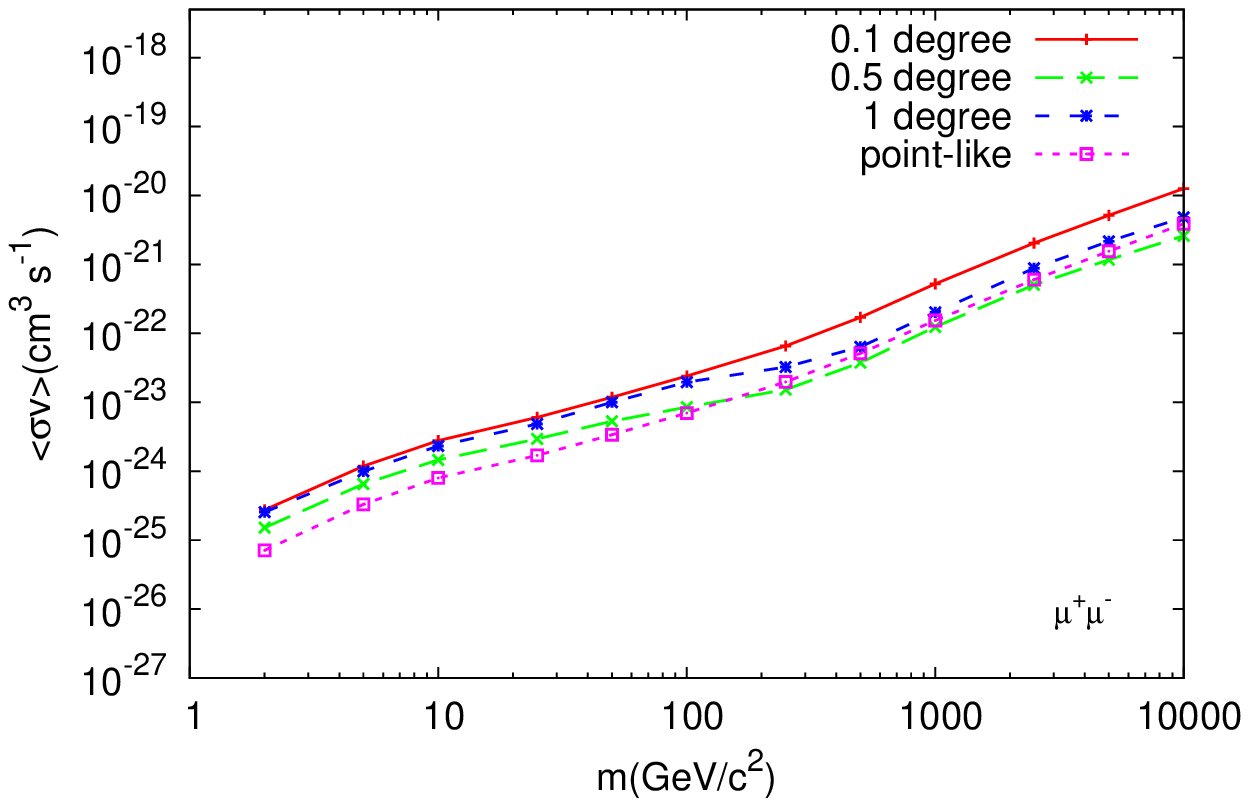}
\includegraphics[width=0.9\columnwidth, angle=0]{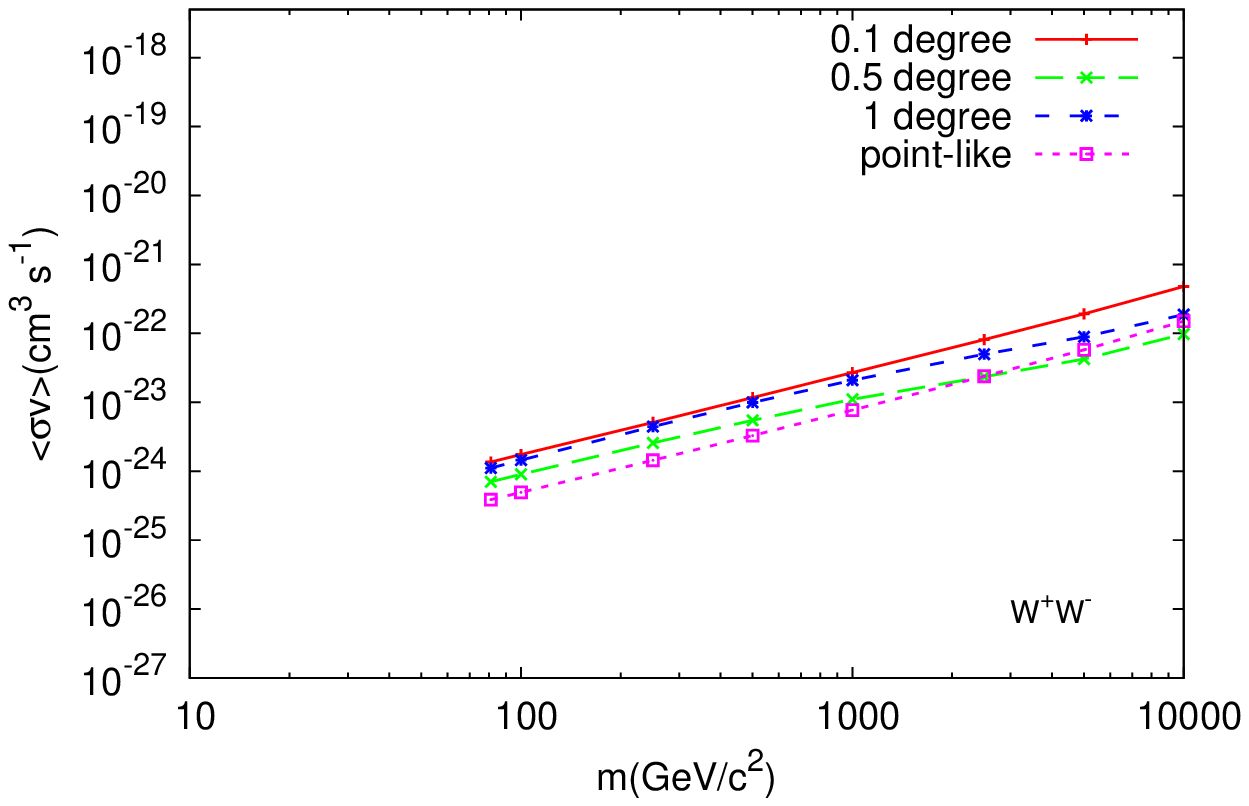}
\figcaption{Constraints on the DM annihilation cross section at $95\%$ CL for six annihilation channels
from about 7 years of Fermi-LAT observation of Reticulum II. This dSph is modeled as a point-like source
and a spatially extended source for three extended angles, respectively.}
\label{fig:zy4}
\end{figure*}

We use the method in Ref. \cite{2013JCAP...03..018S,2016PhRvD..93h3513Z} to derive constraints on DM annihilation from dSph observations.
First, we construct the likelihood map corresponding to the different
integrated fluxes $\Phi(E)$ in each energy bin. For a certain input set of $m_{DM}$, $\left<\sigma v\right>$
and $J$ factor, the combined likelihood in all energy bins for Reticulum II can be estimated as
the following equation
\begin{eqnarray}\label{eq:zy3}
L&&=\prod_{i}L_{i}(\Phi_{i}|D) \nonumber\\
&&\times \frac{1}{\ln(10)J_{obs}\sqrt{2\pi}\sigma}e^{-[\log_{10}(J)-\log_{10}(J_{obs})]^{2}/2\sigma^{2}},
\end{eqnarray}
where the subscript $i$ denotes the $i$-th energy bin, $\Phi_i$ is the $\gamma$-ray signature flux from DM,
and $J_{obs}$ and $\sigma$ are the mean value and deviation of the $J$ factor respectively. In our analysis for Reticulum II, the values of $J_{obs}$ and $\sigma$ are taken to be the mean values and lower uncertainties shown in Table~\ref{tab:zy1}, respectively. For given $m_{DM}$ and $\left<\sigma v\right>$,
the value of $J$ is chosen to make the likelihood value $L$ reach a maximum. Then we obtain a ``cross section-likelihood" table
for Reticulum II. By using this table, we are able to calculate the upper limits on $\left<\sigma v\right>$ at $95\%$ CL,
where the corresponding log-likelihood decreases
by a factor of $2.71/2$ from its maximum \cite{1953Biometrika.40.306,2005NIMPA.551..493R}.
For three typical spatially extended angles, the constraints on the DM annihilation cross section for six
annihilation channels are shown in Fig. \ref{fig:zy4}.
The limits on the DM annihilation cross section for the point source case are stronger than those for the extended source case with a small extended angle $0.1^\circ$ by a factor of $\mathcal{O}(1)$, while they are comparable to those for the extended source case with an extended angle of $0.5^\circ$ or $1^\circ$ at high DM masses.

We also compare our constraints with those set by the Fermi-LAT collaboration in Ref.~\cite{2015ApJ...809L...4D,2017ApJ...834..110A}. We find that the differences are mainly caused by the $J$ factor. For the point-like source case, the $J$ factor of Reticulum II within $0.5^\circ$ is taken to be $\log_{10}J=19.6 \pm 0.7$ in the estimation. This $J$ factor is derived by an optimized Jeans analysis of stellar kinematic data \cite{2015ApJ...808L..36B}. In Ref.~\cite{2015ApJ...809L...4D}, the $J$ factor of $\log_{10}J=19.3 \pm 0.4$ is simply derived by a scaling relation between the distances and spectroscopically determined $J$ factors of known dSphs. In Fig. \ref{fig:zy5}, the red and green lines represent our constraint and that given in Ref.~\cite{2015ApJ...809L...4D} for the $\tau^+\tau^-$ channel, respectively. Although the mean values of the two $J$ factors are different, these two constraints are comparable due to the uncertainties of the $J$ factors. In order to perform a cross-check, we also reproduce the constraint (dark blue line) for the point-like source case with a $J$ factor of $\log_{10}J=19.3 \pm 0.4$ in Fig. \ref{fig:zy5}. We can see that although the analysis details slightly differ, our result is consistent with the Fermi-LAT result for the same $J$ factor.

The Fermi-LAT and DES Collaborations have updated the constraints from the newly discovered dSphs in Ref.~\cite{2017ApJ...834..110A}. In that work, the $J$ factor of Reticulum II is taken to be $\log_{10}J=18.9 \pm 0.6$ within $0.5^\circ$, which is estimated by modeling the velocities through the spherical Jeans equation in Ref.~\cite{2015ApJ...808...95S}. Flat uninformative priors for the DM halo parameters and a constant stellar velocity anisotropy are adopted in that calculation of the $J$ factor, while  different parameterizations of DM and light profiles and anisotropy are used in Ref.~\cite{2015ApJ...808L..36B}. Therefore, the $J$ factors derived from these two groups are different. The constraint on  DM annihilation cross section to $\tau^+\tau^-$ for Reticulum II given by Ref.~\cite{2017ApJ...834..110A} \footnote{http://www-glast.stanford.edu/pub\_data/1203/}
is represented by the pink line in Fig. \ref{fig:zy5}. We can see that this constraint is weaker than the previous one given by the Fermi-LAT collaboration by a factor of $\mathcal{O}(1)$. This discrepancy is mainly caused by the different mean values and uncertainties of the $J$ factors adopted in these works. For a cross-check, we also reproduce the constraint (light blue line) for the point-like source case with the same $J$ factor of $\log_{10}J=18.9 \pm 0.6$ in Fig. \ref{fig:zy5}. We can see that our result is again consistent with the Fermi-LAT result.

\begin{center}
\includegraphics[width=0.9\columnwidth, angle=0]{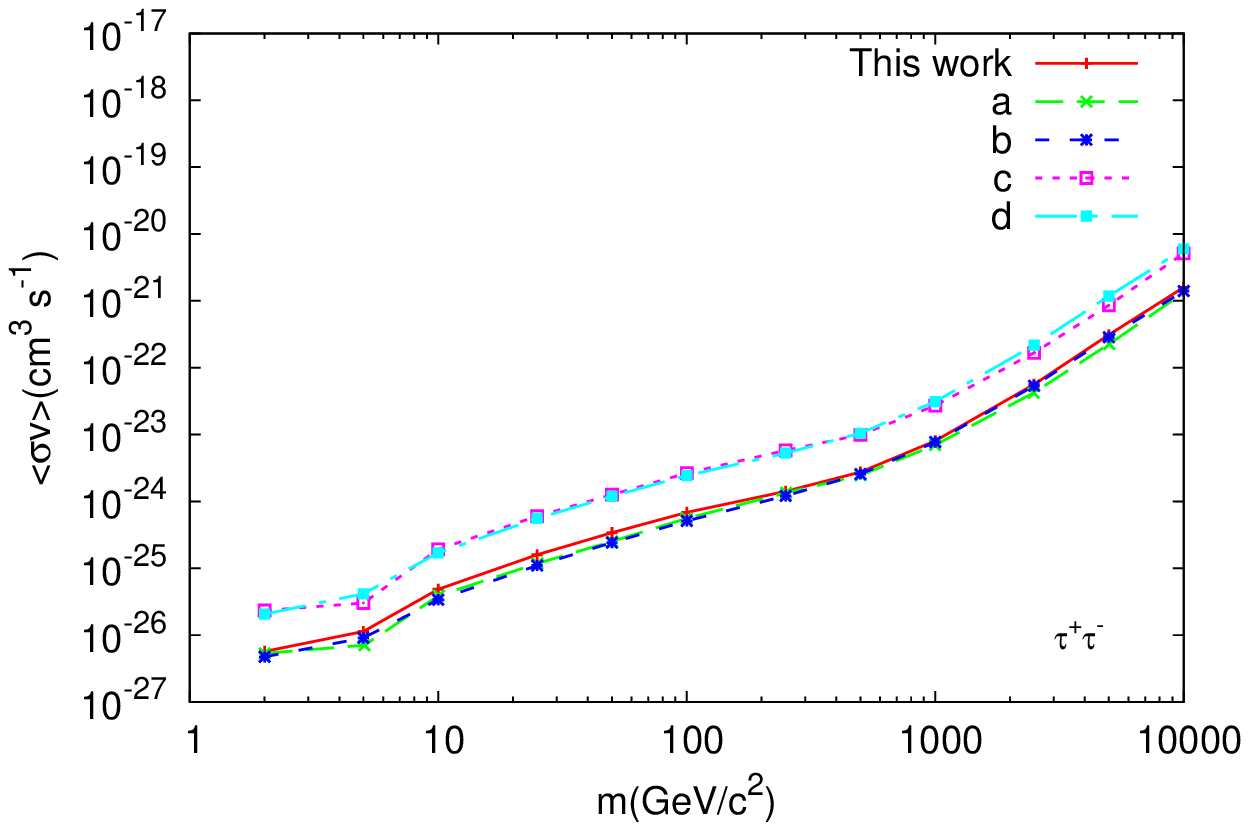}
\figcaption{Upper limits on the DM annihilation cross section at $95\%$ CL for the $\tau^+\tau^-$ channel
for different $J$ factors of Reticulum II. Reticulum II is treated as a point-like source here. The red line represents the constraint set by this work with a $J$ factor of $\log_{10}J=19.6 \pm 0.7$ within $0.5^\circ$. The green (``a") and dark blue (``b") lines represent the constraints set by this work and Ref.~\cite{2015ApJ...809L...4D} with a $J$ factor of $\log_{10}J=19.3 \pm 0.4$, respectively. The pink (``c") and light blue (``d") lines represent the constraints set by this work and Ref.~\cite{2017ApJ...834..110A} with a $J$ factor of $\log_{10}J=18.9 \pm 0.6$, respectively.}
\label{fig:zy5}
\end{center}

\section{Conclusion and discussions}

After the discovery of $16$ new dSphs by the DES,
many studies \cite{2015PhRvL.115h1101G,2015JCAP...09..016H,2015ApJ...809L...4D,2015arXiv151109252L}
have been performed to probe the $\gamma$-ray emission in the
Fermi-LAT data from these candidates. Some studies reported
that there seem a slight excess in Reticulum II and Tuc III,
which are usually treated as point-like sources.
In this work, we probe the $\gamma$-ray emission in the Fermi Pass 8 data from the direction of Reticulum II, which is
assumed to be a spatially extended source. We find that
there is a slight excess at energies between 2.8 GeV and 6.7 GeV with a small TS value like those derived under
the point-like source assumption. An annihilating DM with a mass of $\sim25$ GeV to $\tau^+\tau^-$ final states show
the largest TS value of $\sim 5$-$6$. Different spatially extended angles result in comparable TS values.

Based on above results, we set constraints on the DM annihilation cross section for several channels. We also compare our
constraints under the point-like source assumption with those derived by the Fermi-LAT collaboration. Although the analysis details may slightly differ, our results are consistent with the Fermi-LAT result for the same $J$ factor. We find that the different
$J$ factors derived from different techniques may affect the constraint by a factor of $ \mathcal{O}(1)$.

Finally, it is worth noting that the Fermi-LAT and DES collaborations have probed the $\gamma$-ray emission from all the newly discovered dSphs \cite{2017ApJ...834..110A}.
They found that another two dSphs, namely Indus II and Tuc IV, also have slight excesses,
with a significance of $\sim$2$\sigma$. All the new dSphs are modeled as point-like sources because
the extension properties of their DM halos are not well determined yet. In order to discover the $\gamma$-ray emission from DM annihilation in these dSphs, careful studies on the structures of their DM halos will be necessary.

\acknowledgments{This work is supported by the National Natural Science Foundation of China
under Grants Nos. 11121092, 11033005, 11375202, 11475191, 11475189, by the CAS pilot B program (No. XDB23020000),
and by the National Key Program for Research and Development (No. 2016YFA0400200).}

\end{multicols}

\vspace{10mm}

\begin{multicols}{2}


\end{multicols}

\clearpage

\end{CJK*}
\end{document}